\documentclass[a4paper,11pt]{article}
\usepackage{pos}
 
\title{Recent benchmarks in the Analysis Grand Challenge and integration with Combine (and HS3)}
 
\author[a]{Peter Fackeldey}
\author*[a]{Massimiliano Galli}
\author[b]{Hannah Havel}
\author[c]{Alexander Held}
\author[d]{Iason Krommydas}
\author[e]{Oksana Shadura}
\author[f]{Nick Smith}
 
\affiliation[a]{Princeton University,\\
  Princeton, New Jersey, United States}
\affiliation[b]{Northern Illinois University,\\
  DeKalb, Illinois, United States}
\affiliation[c]{University of Wisconsin-Madison,\\
  Madison, Wisconsin, United States}
\affiliation[d]{Rice University,\\
  Houston, Texas, United States}
\affiliation[e]{University of Nebraska-Lincoln,\\
  Lincoln, Nebraska, United States}
\affiliation[f]{Fermi National Accelerator Laboratory,\\
  Batavia, Illinois, United States}
 
\emailAdd{massimiliano.galli@cern.ch}
 
\abstract{The Analysis Grand Challenge (AGC) showcases an example of HEP analysis. Its
reference implementation uses modern Python packages to realize the main steps, from data
access to statistical model building and fitting. The packages used for data handling and
processing (\texttt{coffea}, \texttt{uproot}, \texttt{awkward-array}) have recently
undergone a series of performance optimizations. While not being part of the HEP Python
(PyHEP) ecosystem, the Combine tool is a pillar of CMS analyses, covering more than 90\% of
the analyses published in the last few years. As such, it is necessary to have Combine
integrated in the PyHEP ecosystem, using the AGC as example. This project also includes, in
the long-term, providing support and integration for the High Energy Physics Statistics
Serialization Standard (HS3), as a way to have a language-independent way of representing
the likelihood and use different frameworks interchangeably. In these proceedings we cover
part of the recent work performed on the AGC and Combine, including: performance
benchmarks, covering benefits introduced by the recent improvements in the data processing
packages; examples of how Combine can be integrated and run in a dedicated infrastructure
(coffea-casa); and examples and plans to integrate HS3 in Combine.}
 
\FullConference{23rd International Workshop on Advanced Computing and Analysis Techniques in Physics Research (ACAT2025)\\
8--12 September 2025\\
Hamburg, Germany\\}
 
\begin{document}
\maketitle
 
\section{The Analysis Grand Challenge}
The IRIS-HEP Analysis Grand Challenge (AGC)~\cite{agc} is designed as a realistic
environment for investigating how analysis methods and tools scale to the demands of the
HL-LHC. Rather than a synthetic benchmark, it is built around a physically meaningful
analysis task: a $t\bar{t}$ cross-section measurement in the single-lepton channel,
including a simple top-quark reconstruction and based on Run~2 CMS Open Data. The task
deliberately exercises all of the workflow aspects encountered in a real analysis, from
data delivery and columnar processing through histogramming to statistical inference.
This end-to-end coverage makes the AGC a natural testbed both for measuring the
performance of the underlying software and for integrating new tools into a common
workflow. In the following we summarise recent benchmarking of the columnar analysis
backends and describe two integration efforts: bringing CMS Combine into the Python
ecosystem, and supporting the HEP Statistics Serialization Standard (HS3) within Combine.
 
\section{Benchmarking the analysis backends}
The Awkward Array library (\texttt{awkward2})~\cite{awkward} has recently undergone
several performance upgrades and now offers different mechanisms to express
\emph{laziness}, the deferral of data loading and computation until the results are
actually required. Two such mechanisms are compared here, both implemented in the
calendar-versioned \texttt{coffea} release line (referred to below as
\texttt{coffea}~202x)~\cite{coffea} and exercised through the AGC task:
 
\begin{itemize}
  \item \textbf{\texttt{dask-awkward}}~\cite{daskawkward} moves laziness to Dask by
  building task graphs and loading data only when the graph is executed. For non-local
  data, this results in a single request over the network. The main costs are that
  building task graphs can be time-consuming for complex analyses, and that some simple
  NumPy-like and machine-learning operations are difficult to express because of the
  \emph{typetracer} arrays used to infer output types and shapes without touching the
  data.
  \item \textbf{Virtual arrays} support all operations and load only the required
  branches, at the cost of issuing multiple network calls when fetching non-local data.
  This behaviour is expected to improve in the near future.
\end{itemize}
 
\begin{figure}[htbp]
  \centering
  \begin{minipage}{0.48\textwidth}
    \centering
    \includegraphics[width=\textwidth]{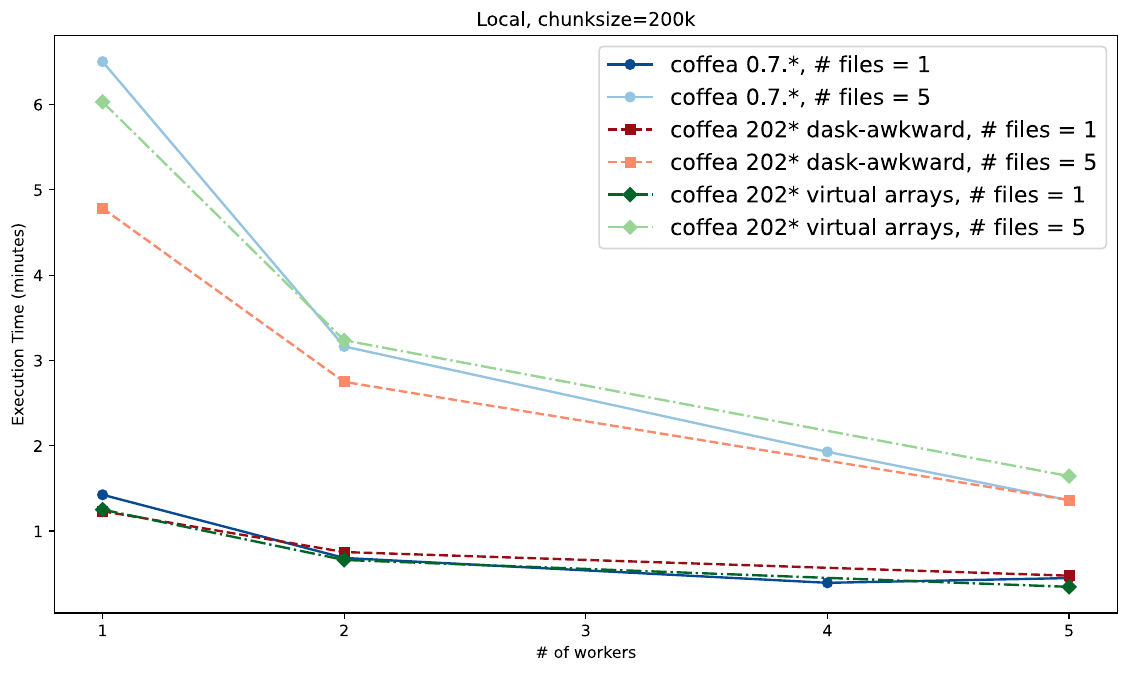}
  \end{minipage}
  \hfill
  \begin{minipage}{0.48\textwidth}
    \centering
    \includegraphics[width=\textwidth]{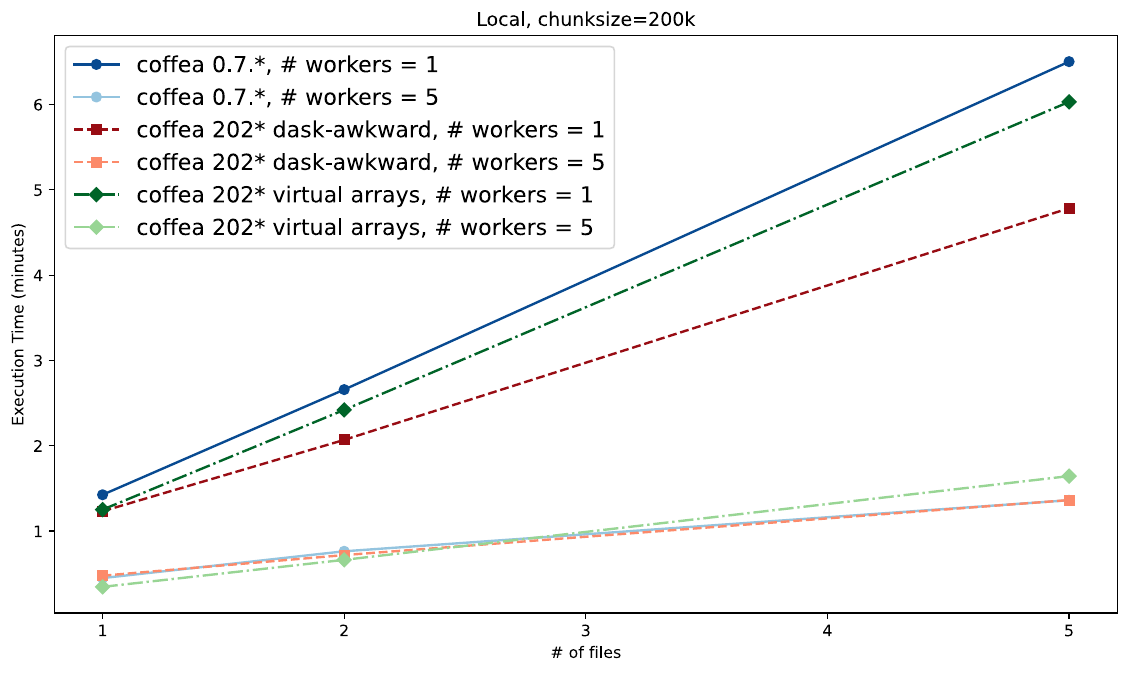}
  \end{minipage}
  \caption{Compute time as a function of number of workers (left) and number of files (right), for data stored locally to the cluster.}
  \label{fig:benchmarks_local}
\end{figure}

\begin{figure}[htbp]
  \centering
  \begin{minipage}{0.48\textwidth}
    \centering
    \includegraphics[width=\textwidth]{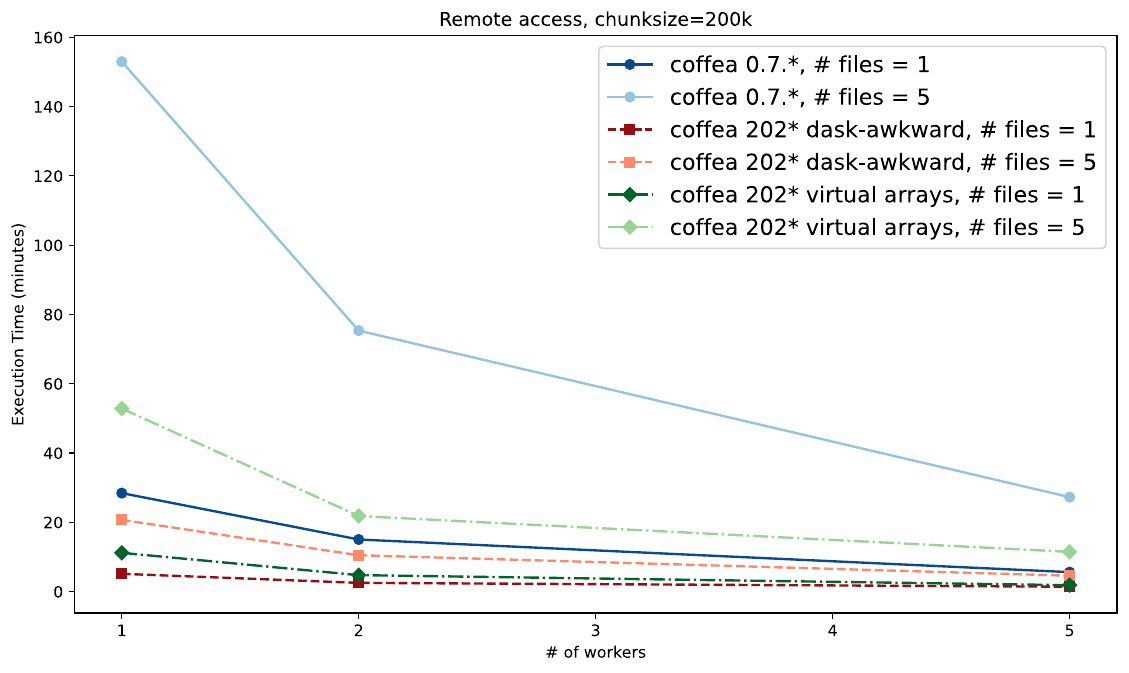}
  \end{minipage}
  \hfill
  \begin{minipage}{0.48\textwidth}
    \centering
    \includegraphics[width=\textwidth]{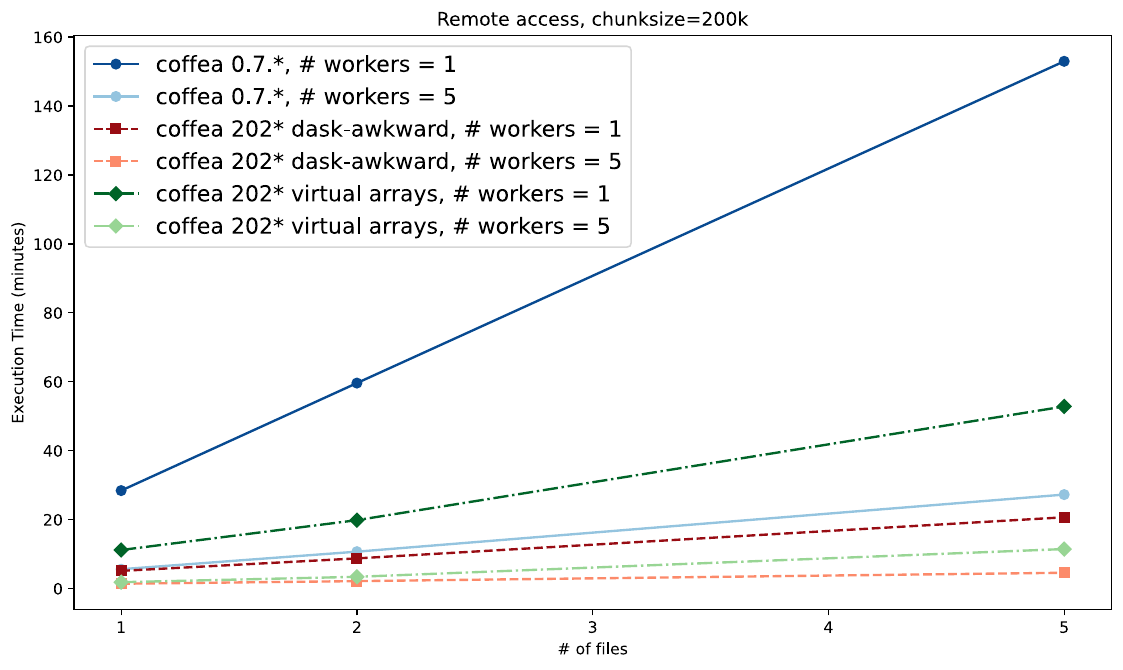}
  \end{minipage}
  \caption{Compute time as a function of number of workers (left) and number of files (right), for data fetched over the network (\texttt{HTTPS} protocol).}
  \label{fig:benchmarks_network}
\end{figure}

Benchmarks were performed by running the AGC analysis task under a controlled setup: the
legacy \texttt{coffea}~0.7 series was compared against \texttt{coffea}~202x with both
\texttt{dask-awkward} and virtual arrays.
Local data access (Figure \ref{fig:benchmarks_local}) was compared with access over the network (Figure \ref{fig:benchmarks_network}).
The compute time was measured as a function of the number of files per sample, the number of workers, and the
chunk size, using the nine samples of the analysis task.

As shown in Figure \ref{fig:benchmarks_local}, for data local to the cluster (no network access), the legacy \texttt{coffea}~0.7 and the new \texttt{coffea}~202x show broadly similar performance.
\texttt{dask-awkward} performs best when using a single worker, benefiting from task-graph optimisation; the differences between backends largely disappear as the number of files and workers increases, i.e.\ in the more realistic operating regime.
Tests that vary the chunk size highlight the reduction in Python overhead achieved in \texttt{awkward2}.
 
For remotely located samples accessed over \texttt{HTTPS} (Figure \ref{fig:benchmarks_network}), \texttt{coffea}~202x benefits from improvements in \texttt{uproot5}~\cite{uproot} and outperforms the legacy series. In this regime \texttt{dask-awkward} performs better than virtual arrays, because it issues a
single call to the network whereas virtual arrays currently issue multiple calls;
reducing this overhead is work in progress.
 
 
%
%
%
 
\section{Integration with CMS Combine}
CMS Combine~\cite{combine} is a pillar of CMS analyses, covering more than 90\% of the
analyses published in the last few years. Although it is not part of the PyHEP ecosystem,
its central role makes integration into that ecosystem highly desirable, and here we again
use the AGC as a concrete example. Installation has been considerably simplified by
packaging Combine on \texttt{conda-forge}, so that a working installation is obtained with
a single command:
 
\begin{verbatim}
$ conda install conda-forge::cms-combine
\end{verbatim}
 
\noindent Implementing the statistical-inference step of the AGC task in Combine is useful
both for teaching purposes and for cross-checks against other fitting frameworks such as
\texttt{pyhf}~\cite{pyhf}.
Beyond a local installation,
Combine can also be run on dedicated analysis infrastructure such as
coffea-casa~\cite{coffeacasa}, demonstrating that the full AGC workflow, including the
Combine-based inference step, can be executed on a shared analysis facility.
 
 
\section{HS3 support in Combine}
The HEP Statistics Serialization Standard (HS3)~\cite{hs3} is a proposed standard for
describing statistical models, procedures and results in high-energy physics using human-
and machine-readable representations based on JSON files. Work is ongoing to support HS3
in CMS Combine through a ``hybrid'' approach, in which a Combine datacard can point to an
HS3 JSON file rather than encoding all shape information itself. Concretely, the
\texttt{shapes} lines of a datacard that would normally reference ROOT files can instead
reference the corresponding HS3 JSON file:
 
\begin{verbatim}
shapes *     * all_histograms_fps4_$CHANNEL.json $PROCESS $PROCESS_$SYSTEMATIC
shapes ttbar * all_histograms_fps4_$CHANNEL.json ttbar   ttbar_$SYSTEMATIC
\end{verbatim}
 
\noindent The AGC again serves as a convenient example on which to test and validate the
implementation.
 
\section{Summary and outlook}
The IRIS-HEP Analysis Grand Challenge continues to be extended and broadened. Benchmarks
have been run to quantify the performance improvements of the backbone packages of the
PyHEP ecosystem, and the challenge is being integrated with packages beyond that
ecosystem, notably CMS Combine and HS3. Looking ahead, the AGC is evolving into an
``integration challenge'' that will incorporate more facilities and data, and more complex
tasks closer to a full-scale analysis, together with additional analysis examples.
 
\section*{Acknowledgments}
This work was supported by the U.S.\ National Science Foundation (NSF) under cooperative
agreements OAC-1836650 and PHY-2323298 (IRIS-HEP).
 
%
\bibliographystyle{JHEP}
\bibliography{refs}
 
\end{document}